\documentclass[prl,twocolumn,showpacs,preprintnumbers,amsmath,amssymb]{revtex4}
\usepackage{graphicx}
\usepackage{latexsym}
\usepackage{color}

\newcommand{\be}{\begin{equation}}
\newcommand{\ee}{\end{equation}}

\def\ie{{\it i.e.}\ }

\def\rms{{\it r.m.s.}\ }

\def\al{Alfv\'en\ }

 \newcommand{\bB}{\mathbf{B}}

\newcommand{\alp}{\alpha}

\newcommand{\bj}{\mathbf{j}}

 \newcommand{\bv}{\mathbf{v}}
 
\newcommand{\bus}{\mathbf{u_s}} \newcommand{\bBs}{\mathbf{B_s}}

\begin{document}
\title{Numerical study of dynamo action at low magnetic Prandtl numbers}

\author{Y. Ponty$^1$, P.D. Mininni$^2$, D.C. Montgomery$^3$, J.-F. Pinton$^4$, 
H. Politano$^1$ and A. Pouquet$^2$}
\affiliation{$^1$ CNRS UMR6202, Laboratoire Cassiop\'ee, Observatoire de la C\^ote d'Azur, 
BP 4229, Nice Cedex 04, France \\
$^2$ASP/NCAR, P.O. Box 3000, Boulder, Colorado 80307-3000, U.S.A. \\
$^3$Dept. of Physics and Astronomy, Dartmouth College, 
Hanover, NH 03755, U.S.A. \\
$^4$CNRS UMR5672, Laboratoire de Physique, \'Ecole Normale 
Sup\'erieure de Lyon, 46 All\'ee d'Italie, 69007 Lyon, France}

\begin{abstract}
We present a three--pronged numerical approach to the dynamo problem 
at low magnetic Prandtl numbers $P_M$. The difficulty of resolving 
a large range of scales is circumvented by combining Direct Numerical 
Simulations, a Lagrangian-averaged model, and Large-Eddy Simulations (LES). 
The flow is generated by the Taylor-Green forcing; it combines a well 
defined structure at large scales and turbulent fluctuations at small 
scales. Our main findings are:
(i) dynamos are observed from $P_M=1$ down to $P_M=10^{-2}$; 
(ii) the critical magnetic Reynolds number increases sharply with $P_M^{-1}$ 
as turbulence sets in and then saturates; 
(iii) in the linear growth phase, the most unstable magnetic modes move 
to small scales as $P_M$ is decreased and a Kazantsev $k^{3/2}$ spectrum 
develops; then the dynamo grows at large scales and modifies the turbulent 
velocity fluctuations.
\end{abstract}
\maketitle

The generation of magnetic fields in celestial bodies occurs 
in media  for which the viscosity $\nu$ and the magnetic diffusivity $\eta$ 
are vastly different. For example, in the interstellar medium the magnetic 
Prandtl number $P_M=\nu/\eta$ has been estimated to be as large as $10^{14}$, 
whereas in stars such as the Sun and for planets such as the Earth, it can 
be very low ($P_M<10^{-5}$, the value for the Earth's iron core). Similarly 
in liquid breeder reactors and in laboratory experiments in liquid metals, 
$P_M\ll 1$. At the same time, the Reynolds number $R_V=UL/\nu$ 
($U$ is the \rms velocity, $L$ is the integral scale of the flow) is very large, and the 
flow is highly complex and turbulent, with prevailing non-linear effects 
rendering the problem difficult to address. If in the smallest scales of 
astrophysical objects plasma effects may prevail, the large scales are 
adequately described by the equations of magnetohydrodynamics (MHD), 
\begin{eqnarray}
&&\frac{\partial {\bf v}}{\partial t}+ {\bf v} \cdot \nabla {\bf v} =
-\nabla {\cal P} + \bj \times \bB +\nu \nabla^2 \bv + {\bf F} \label{E_MHDv} \\
&&\frac{\partial {\bf B}}{\partial t}+ {\bf v} \cdot \nabla {\bf B} = 
\bB \cdot \nabla {\bf v}
+\eta \nabla^2 {\bf B} \ , 
\label{E_MHDb}  
\end{eqnarray}
together with ${\bf \nabla} \cdot {\bf v} =0$, $\nabla \cdot {\bf B} =0$, and assuming a constant mass density.  Here, $\bv$ is the velocity field normalized to the \rms fluid flow speed,  and $\bB$ the magnetic field converted to velocity units by means of an 
equivalent \al speed. ${\cal P}$ is the pressure and $\bj=\nabla \times \bB$ 
the current density. ${\bf F}$ is a forcing term, responsible for the generation of the flow (buoyancy and Coriolis in planets, mechanical drive in experiments).

Several mechanisms have been studied for dynamo 
action, both analytically and numerically, involving in particular 
the role of helicity~\cite{moffat} (\ie the correlation 
between velocity and its curl, the vorticity) for dynamo growth at 
scales larger than that of the velocity, and the role of chaotic 
fields for small-scale growth of magnetic excitation (for a  
recent review, see~\cite{axel}). Granted that the stretching 
and folding of magnetic field lines by velocity 
gradients overcome dissipation, dynamo action takes place above 
a critical magnetic Reynolds number  $R_M^c$, with $R_M= P_M R_V = UL/\eta$. 
Dynamo experiments engineering constrained helical flows of liquid sodium 
have been successful~\cite{KarlsruherigaGailitisStefaniTilgner}.
However, these experimental setups do not allow for a complete 
investigation of the dynamical regime, and many groups have 
searched to implement unconstrained dynamos~\cite{GydroSpecialIssue}. 
Two difficulties arise: first, turbulence now becomes fully developed 
with velocity fluctuations reaching up to 40\% of the mean; second, 
it is difficult to engineer flows with helical small scales so that 
the net effect of turbulence is uncertain. Recent Direct Numerical 
Simulations (DNS) address the case of randomly forced, 
non-helical flows with magnetic Prandtl numbers from 1 to 0.1. 
Contradictory results are obtained: it is shown in~\cite{alex04} 
that dynamo action can be inhibited for $P_M<1/4$, 
while it is observed in~\cite{axel} that the dynamo threshold 
increases as $P_M^{-1/2}$ down to $P_M \sim 0.3$. Experiments  
made in von K\'arm\'an geometries (either spherical or cylindrical) 
 have reached $R_M$ values up to 60~\cite{pefbour}. Also, MHD turbulence at 
low $P_M$ has been studied in the idealized context of turbulent 
closures \cite{kn67}. In this context, turbulent dynamos are found, 
and the dependences of $R_M^c$ 
upon three quantities are studied, namely $P_M$, the relative rate of helicity 
injection, and the forcing scale. An increase of $\sim 20\%$ in $R_M^c$ 
is observed as $P_M$ decreases from 1 to $\sim 3 \times 10^{-5}$. 
Recently, the Kazantsev-Kraichnan~\cite{kazan} 
model of $\delta$-correlated velocity fluctuations has been used to study the effect of 
turbulence. 
It is shown that the threshold increases with the 
rugosity of the flow field~\cite{dario02}, and that turbulence can 
either increase or decrease the dynamo threshold depending on the fine 
structure of the velocity fluctuations~\cite{leprovost}.

There is therefore a strong motivation to study how the dynamo threshold varies as $P_M$ is progressively decreased, for a given flow. In  this letter we focus on a situation where the flow forcing is not random,  but generates a well defined geometry at large scales, with turbulence developing naturally at small scales as the $R_V$ increases. This 
situation complements recent numerical  works~\cite{alex04,axel,dario02,leprovost} and is quite relevant for  planetary and laboratory flows. Specifically, we consider the swirling 
flow resulting from the Taylor-Green forcing~\cite{meb}:
\begin{equation} 
{{\bf  F}_{\rm TG}(k_0)}= { 2F } \,  \left[ 
\begin{array}{c} 
\sin(k_0~x) \cos(k_0~y) \cos(k_0~z) \\ 
- \cos(k_0~x) \sin(k_0~y) \cos(k_0~z)\\ 0  
\end{array} \right]  \ ,
\label{eq:Ftg}
\end{equation} 
with $k_0=2$, so that dynamo action is free to develop at scales  larger or smaller than the forcing scale $k_f=k_0 \sqrt{3}$. This  force generates flow cells that have locally differential rotation  and helicity, two key ingredients for dynamo action~\cite{moffat,axel}. Note that the net helicity, {\it i.e.} averaged in time and space, is zero in the $2\pi$-periodic domain. However strong local fluctuations of helicity are always present in the flow. Small scales are statistically non-helical. The resulting flow also shares 
similarities with the Maryland, Cadarache and Wisconsin sodium experiments~\cite{GydroSpecialIssue}, and it has motivated several numerical studies at $P_M \sim 1$~\cite{NoreDuddleyJames,MarieBourgoin}. 

\begin{table}[h!]    
\begin{tabular}{|c|c|c|c|c|c|c|c|c|}\hline
code          & $N$                 & $R_V$       & $L$ & $R_M^c$   & $1/P_M^c$      & $k_{\rm MAX}$ & $k_D$ & $\rho$ \\ \hline\hline
 {DNS}      &  {$64$} &  {30.5}  &  {2.15}      &  {28.8}   &  {1.06}         &       2      	    &        5   &     -7.2 \\ \hline
 {DNS}      &  {$64$} &  {40.5}   &  {2.02}     &  {31.7}   &  {1.28}         &      2                 &       5    &     -6.3  \\ \hline
 {DNS}      &  {$64$} &  {128}    &  {1.9}       &  {62.5}    &  {2.05}        &      4                 &       9     &     -3.5  \\ \hline
 {DNS}      &  {$128$}&  {275}    &  {1.63}     &  {107.9}  &  {2.55}       &      5                  &     11      &    -2.15  \\ \hline
 {DNS}      &  {$256$}&  {675}    &  {1.35}     &  {226.4}  &  {2.98}       &      7                  &     21     &    $\sim -5/3$  \\ \hline
 {DNS}      &  {$512$}&  {874.3}  &  {1.31}     &  {192.6}  &  {4.54}       &      9                  &     26     &    $\sim -5/3$  \\ \hline\hline
 {LAMHD}   &  {$64$} &  {280}     &  {1.68}     &  {117.3} &  {2.38}        &     6                   &    11      &     -2.25 \\ \hline
 {LAMHD}   &  {$128$}&  {678.3} &  {1.35}     &  {256.6} &  {2.64}         &     8                   &    12      &    $\sim -5/3$  \\ \hline
 {LAMHD}   &  {$128$}&  {880.6}  &  {1.32}    &  {242.1}   &  {3.64}       &     9                   &     22     &    $\sim -5/3$  \\ \hline
 {LAMHD}   &  {$256$}&  {1301.1}&  {1.3}     &  {249.3}  &  {5.22}       &     9                  &     31       &   $\sim -5/3$  \\ \hline
 {LAMHD}   &  {$512$}&  {3052.3}&  {1.22}     &  {276.4} &  {11.05}      &     10                 &     45       &  $\sim -5/3$  \\ \hline\hline
 {LES}       &  {$128$}&  {2236.3} &  {1.37}   &  {151.9} &  {14.72}     &      5                &      21       &    $-5/3$  \\ \hline
 {LES}       &  {$256$}&  {5439.2} &  {1.39}   &  {141}   &  {38.57}     &      5                &      31       &    $-5/3$  \\ \hline
 {LES}       &  {$512$}&  {12550}  &  {1.42}   &  {154.6} &  {81.19}     &      5                &      40       &    $-5/3$  \\ \hline\hline
\end{tabular}
\caption{Parameters of the computation: code used, linear grid resolution $N$, Reynolds number $R_V$, integral  length scale $L$
(defined from the kinetic energy spectrum $L = 2\pi \int{k^{-1} E_V(k) dk}/ \int{E_V(k) dk}$), critical magnetic Reynolds number $R_M^c$, inverse  magnetic Prandtl number $1/P_M^c$, wavenumber $k_{\rm MAX}$ with the largest magnetic energy, characteristic wavenumber $k_D$ of  magnetic field gradients (defined as the maximum of the current density spectrum), and kinetic spectral index $\rho$ in the range $[k_{\rm MAX}, k_D]$. The values of $\rho$, $L$ and $U$ used in the definitions of the Reynolds and magnetic Prandtl numbers, are computed as time averages during the steady state of the hydrodynamic simulation; $k_{MAX}$ and $k_D$ are computed as time averages during the linear regime of the dynamo simulation closest to $R_M^c$.}
\label{tab1}  
\end{table} 

\begin{figure}[b!]
\centerline{\includegraphics[width=8.7cm]{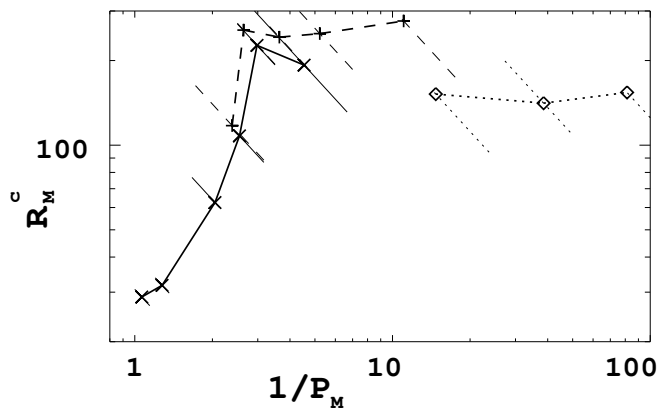}}
\caption{$R_M^c$ for dynamo action versus inverse 
         $P_M$. Symbols are: $\times$ (DNS), $+$ (LAMHD), and $\diamond$ 
         (LES). Transverse lines indicate error bars in the 
         determination of $R_M^c$, as the distance between growing 
         and decaying runs at a constant $R_V$.}
\label{growth} 
\end{figure}

Our numerical study begins with DNS in a 3D periodic domain.  The code 
uses a pseudo-spectral algorithm, an explicit second order Runge-Kutta 
advance in time, and a classical dealiasing rule --- the last resolved 
wavenumber is $k=N/3$ where $N$ is the number of grid points per 
dimension. Resolutions from $64^3$ to $512^3$ grid points are used, 
to cover $P_M$ from 1 to $1/5$. However, DNS are limited in the 
Reynolds numbers and the (lowest) $P_M$ they can reach. We then use a 
second method, the LAMHD (or $\alpha$) model, in which we integrate the 
Lagrangian-averaged MHD equations \cite{LAMHD,mmp04}. This formulation 
leads to a drastic reduction in the degrees of freedom at small 
scales by the introduction of smoothing lengths $\alpha_V$ and $\alpha_M$. 
The fields are written as the sum of filtered (smoothed) and fluctuating 
components: $\bv = \bus + \delta \bv$, $\bB = \bBs + \delta \bB$, with 
$\bus = G_{\alpha_V} \otimes \bv$, $\bBs = G_{\alpha_M} \otimes \bB$, 
where `$\otimes$' stands for convolution and $G_\alpha$ is the smoothing 
kernel at scale $\alpha$, 
$G_{\alpha}({\bf r}, t) = \exp [-r/\alpha]/4\pi \alpha^2 r$. Inversely, 
the rough fields can be written in terms of their filtered counterparts 
as: $\bv = (1-\alp_V^2 \nabla^2)\ \bus$ and 
$\bB = (1-\alp_M^2 \nabla^2)\ \bBs$. In the resulting equations, 
the velocity and magnetic field are smoothed, but not the fields' sources, 
\ie the vorticity and the current density~\cite{mp02}. This model 
has been checked in the fluid case against experiments and DNS of 
the Navier-Stokes equations in 3D~\cite{CFHOTW99}, as 
well as in MHD in 2D \cite{mmp04}. Finally, in order to reach still 
lower $P_M$, we implement an LES model. LES 
are commonly used and well tested in fluid dynamics against laboratory 
experiments and DNS in a variety of flow configurations \cite{parviz}, 
but their extension to MHD is still in its infancy (see however 
\cite{LESMHD}). We use a scheme as introduced in~\cite{ppp04}, aimed at 
integrating the primitive MHD equations with a turbulent velocity field 
all the way down to the magnetic diffusion with no modeling in the induction 
equation but with the help of a dynamical eddy viscosity ~\cite{CL81}:
\begin{equation} 
\nu(k,t) = 0.27 [1 + 3.58 (k/K_c)^8] \sqrt{E_V(K_c,t)/K_c} \ ;
\label{eq:chollet-lesieur} 
\end{equation} 
$K_c$ is the cut-off wavenumber of the velocity field, and $E_V(k,t)$ is 
the one-dimensional kinetic energy spectrum. A consistency condition for 
our approach is that the magnetic field fluctuations be fully resolved 
when $2\pi/K_c$ is smaller than the magnetic diffusive scale  
$\ell_\eta \sim L/R_M^{3/4}$.

The numerical methods, parameters of the runs, and associated characteristic quantities are given in Table I. In all cases, we first perform a hydrodynamic run, lasting about 10 turnover times, to obtain a statistically steady flow. Then we add a seed magnetic field, and monitor the growth of the magnetic energy $E_M$ for a time that depends on the run resolution; it is of the order of 1 magnetic diffusion time $\tau_{\eta}=(2\pi)^2/\eta$ at $64^3$, but it drops down to $\tau_{\eta}/5$ at $512^3$. We define the magnetic 
energy growth rate as $\sigma = d\log E_M/dt$, computed in the linear regime ($t$ is in units of large scale turnover time). The dynamo  threshold corresponds to $\sigma = 0$. For each configuration (Table I), we make several MHD simulations with different $P_M$, varying $\eta$, and for a fixed $R_V$ defined by the hydrodynamic run. We bound the marginal growth between clearly decaying and growing evolutions of the magnetic energy. This procedure is unavoidable because of the critical slowing down near 
threshold.

\begin{figure}[t!]
\includegraphics[width=8.7cm]{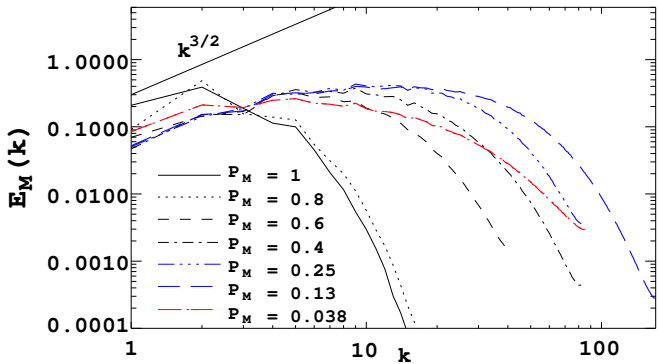}
\caption{Magnetic spectra for $P_M=1$ to $P_M=0.4$ (DNS),  $P_M=0.25,0.13$ (LAMHD), 
         $P_M=0.038$ (LES), at a time within 
         the linear growth of magnetic energy. 
}
\label{lowR}
\end{figure}

At $P_M=1$, the dynamo self-generates at $R_M^c=30$. As $P_M$ is lowered, 
we observe in the DNS that the threshold reaches 
$R_M^c=70$ at $P_M=0.5$ and then increases steeply to $R_M^c=226$ at 
$P_M=0.3$; at lower $P_M$ it does not increase anymore, but drops 
slightly to a value of 200 at $P_M=0.2$ (Fig.1 and Table I). We 
then continue with LAMHD simulations to reach lower $P_M$. To ensure 
the consistency of the method, we have run overlapping 
DNS and LAMHD simulations in the range from $P_M=0.4 - 0.2$, the agreement 
of the two methods being evaluated by the matching of the magnetic energy 
growth (or decay) rates for identical $(P_M, R_M)$ parameters. We have 
observed that a good agreement between the two methods can be reached if 
one uses two different filtering scales $\alp_V$ and $\alp_M$ in LAMHD, 
chosen to maintain a dimensional relationship between the 
magnetic and kinetic dissipation scales, namely $\alp_V/\alp_M = P_M^{3/4}$. 
Our observation with the LAMHD computations is that the steep increase in 
$R_M^c$ to a value over 250 is being followed by a plateau for $P_M$ 
values down to 0.09. We do note a small but systematic trend of the 
LAMHD simulations to overestimate the threshold compared to DNS. We 
attribute it to the increased turbulent intermittency generated by the 
$\alpha$ model, but further investigations are required to describe 
fully this effect. The LES simulations allow us to further our 
investigation;  with this model the threshold for dynamo self-generation 
remains constant, of the order of 150, for $P_M$ between $10^{-1}$ and 
$10^{-2}$. 

In regards to the generation of dynamo action in the Taylor-Green geometry we thus find: (i) at all $P_M$ investigated a dynamo threshold exists; (ii) as $P_M$ drops below 0.2 - 0.3, the critical $R_M^c$ levels and remains of the order of 200; (iii) the steep initial increase in $R_M^c$ is identified with the development of an inertial range in the spectra of kinetic energy. As the kinetic energy spectrum grows progressively into a Kolmogorov $k^{-5/3}$  spectrum, $R_M^c$ ceases to have significant changes -- cf. Table~\ref{tab1}. 

\begin{figure}[t!]
 \includegraphics[width=8cm]{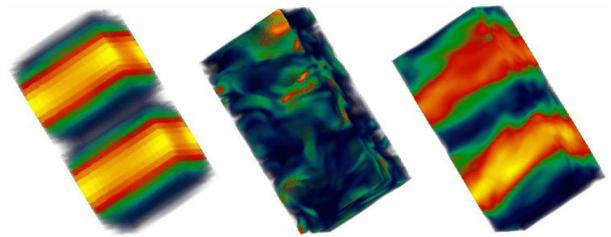}
\caption{Spatial distributions of the magnetic energy 
         for two Taylor-Green cells (DNS) : $P_M=1$, $R_V \sim 30$ at $t=20$
         (left), $P_M=0.4$, $R_V=675$ at $t=20$ (center), and 
          $t=150$ (right).} 
\label{highR} \end{figure}

We plot in Fig.~\ref{lowR} the magnetic energy spectra $E_M(k)$ during 
the linear growth phase, at identical instants when normalized by the 
growth rate.  Four features are noteworthy: first, the dynamo grows from 
a broad range of modes;  second, the maximum of  $E_M(k)$  moves 
progressively to smaller scales as $P_M$ decreases, a result 
already found numerically in \cite{axel};  third, a self-similar 
magnetic spectrum, $E_M(k)\sim k^{3/2}$,  develops at the beginning 
during the linear growth phase --- as predicted by Kazantsev~\cite{kazan} 
and found in other numerical simulations of dynamo generation by turbulent 
fluctuations~\cite{axel,alex04}. This is a feature that thus persists when 
the flow has well defined mean geometry in addition to turbulence. Lastly 
we observe that the initial magnetic growth at small scales is 
always followed by a second phase where the magnetic field grows in the 
(large) scales of the Taylor-Green flow.
Figure~3 shows renderings of the magnetic energy and compare 
low and high Reynolds number cases. When the dynamo is generated 
at low Reynolds number ($R_V \sim 30$ and $P_M=1$), the magnetic field is smooth. As 
$P_M$ decreases and the dynamo grows from a turbulent field, one first 
observes a  complex  magnetic field pattern -- for $t<40$, in the 
example shown in Fig.3(center). But as non-linear effects develop (here for 
times $t>40$) a large scale mode ($k=2$) dominates the 
growth with a structure that is similar to the one at low 
$R_V$. The initial growth of small scale magnetic fields and the 
subsequent transfer to a large scale dynamo mode is also clearly visible 
on the development in time of the magnetic and kinetic energies, 
in a high $R_V$ case, as shown in Fig.~4. During the linear 
growth, a wide interval of modes increase in a self-similar fashion, 
accounting for the complexity of the dynamo field - cf. Fig.~3(center). At 
a later time, the large scale field grows and the 
kinetic energy spectrum $E_V(k)$ is progressively modified at inertial scales. 
The spectral slope changes from a Kolmogorov $k^{-5/3}$ scaling to a 
steeper, close to $k^{-3}$, regime~\cite{kmoins3}. The effect is to 
modify the turbulent scales and to favor the dynamo mode that is allowed 
by the large scale flow geometry. This is consistent with the development 
of a $k^{-5}$ magnetic spectrum, observed in the 
Karlsruhe dynamo experiment~\cite{muller}. It also corroborates the claim 
\cite{petrelis} that the saturation of the turbulent dynamo starts with  
the back-reaction of the Lorentz force on the turbulent fluctuations.  

\begin{figure}[t!]
\includegraphics[width=8.7cm]{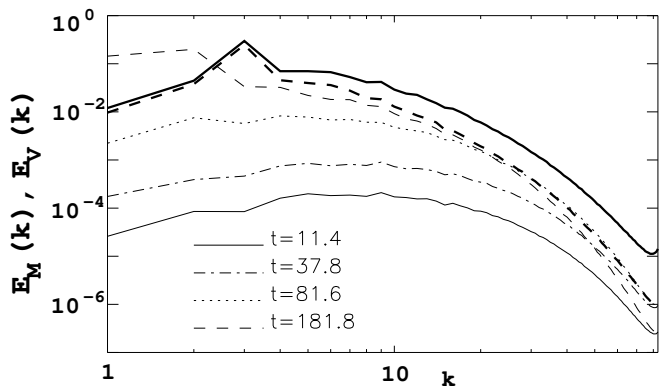}
\caption{Magnetic (thin line) and kinetic (thick line) spectra 
            as a function of time at $P_M=0.4$ (DNS).}
\label{specs}
\end{figure}

To conclude, using a combination of DNS, LAMHD modeling and LES, we 
show that, for the Taylor-Green flow forcing, there is a strong increase 
in the critical magnetic Reynolds number for dynamo action when $P_M$ 
is decreased, directly linked to the development of turbulence; and 
it is followed by a plateau on a large range of $P_M$ from $\sim 0.25$ 
to $\sim 10^{-2}$.  In a situation with both a mean flow and turbulent 
fluctuations, we find that the selection of the dynamo mode results 
from a subtle interaction between the large and small scales.

{\bf Acknowledgements}
\ We thank D. Holm for discussions about the $\alpha$ model, and H. Tufo for
providing computer time at UC-Boulder, NSF ARI grant CDA--9601817. NSF grants 
ATM--0327533 (Dartmouth) and CMG--0327888 (NCAR) are acknowledged. 
JFP, HP and YP thank CNRS Dynamo GdR, and INSU/PNST and PCMI Programs 
for support. Computer time was provided by NCAR, PSC, NERSC, and IDRIS (CNRS).


\begin{thebibliography}{aa}

\bibitem{moffat}
H.K.~Moffatt. {\it Magnetic Field Generation in Electrically Conducting Fluids}, (Cambridge U.P., Cambridge, 1978).

\bibitem{axel}
A. Brandenburg and K. Subramanian, astro-ph/0405052, submitted to {\it Phys. Rep.} (2004).

\bibitem{KarlsruherigaGailitisStefaniTilgner}
A. Gailitis, 
{\it Magnetohydrodynamics}, {\bf 1}, 63 (1996).
A. Tilgner, 
{\it Phys. Rev. A}, {\bf 226}, 75 (1997).
R.~Steglitz and U.~M{\"u}ller,  
{\it Phys. Fluids}, {\bf 13}(3), 561 (2001).
A.~Gailitis, et~al., 
{\it Phys. Rev. Lett.}, {\bf 84}, 4365 (2000).

\bibitem{GydroSpecialIssue}
See ``MHD dynamo  experiments'', special issue of  {\it Magnetohydodynamics}, {\bf 38}, (2002).

\bibitem{alex04}
A. Schekochihin et al., 
{\it New J. Physics} {\bf 4}, 84 (2002);
A. Schekochihin et al. 
{\it Phys. Rev. Lett.} {\bf 92}, 054502 (2004).

\bibitem{pefbour}
N.L. Peffley, A.B. Cawthrone, and D.P. Lathrop,
{\it Phys. Rev. E}, {\bf 61}, 5287 (2000).
M. Bourgoin et al.
{\it Physics of Fluids}, {\bf 14}(9), 3046 (2001).

\bibitem{kn67}
R.H. Kraichnan and S. Nagarajan, {\it Phys. Fluids} {\bf 10}, 859 (1967);
J. L\'eorat, A. Pouquet, and U. Frisch,
{\it J. Fluid Mech.}, {\bf 104}, 419 (1981).

\bibitem{kazan}
A.P. Kazantsev, {\it Sov. Phys. JETP} {\bf 26}, 1031  (1968);
R.H. Kraichnan, {\it Phys. Fluids} {\bf 11}, 945 (1968).

\bibitem{dario02}
S. Boldyrev and F. Cattaneo, {\it Phys. Rev. Lett.}, {\bf 92}, 144501 (2004); 
D. Vincenzi, {\it J. Stat. Phys.} {\bf 106}, 1073 (2002).

\bibitem{leprovost}
N. Leprovost and B. Dubrulle, astro-ph/0404108, (2004).

\bibitem{meb}
M. Brachet,
{\it C. R. Acad. Sci. Paris} {\bf 311}, 775 (1990).

\bibitem{NoreDuddleyJames}
N.L. Dudley and  R.W. James,
 {\it Proc. Roy. Soc. Lond.}, {\bf A425}, 407 (1989).
C. Nore et al. 
{\it Phys. Plasmas}, {\bf 4},1 (1997).

\bibitem{MarieBourgoin}
L. Mari\'e et al., 
{\it Eur. J. Phys. B}, {\bf 33}, 469 (2003).
M. Bourgoin et al., 
{\it Phys. Fluids}, {\bf 16}, 2529 (2004).

\bibitem{LAMHD}
D.D. Holm, {\it Physica D} {\bf 170}, 253 (2002);
{\it Chaos} {\bf 12}, 518 (2002).

\bibitem{mmp04}
P.D. Mininni, D.C. Montgomery, and A. Pouquet , submitted to {\it Phys. Fluids}.

\bibitem{mp02}
D.C. Montgomery and A. Pouquet, {\it Phys. Fluids} {\bf 14}, 3365(2002).

\bibitem{CFHOTW99}
S.Y. Chen et al., 
{\it Phys.  Fluids}  {\bf 11}, 2343 (1999);
S.Y. Chen et al., 
{\it Physica D} {\bf 133}, 66 (1999).


\bibitem{parviz}
R.S. Rogallo and P. Moin, {\it Ann. Rev. Fluid Mech.} {\bf  16}, 99 (1984);
C. Meneveau and J. Katz, {\it Ann. Rev. Fluid Mech.} {\bf 32}, 1 (2000).

\bibitem{LESMHD}
A. Pouquet, J.  L\'eorat, and U. Frisch,  {\it J. Fluid Mech.}, {\bf 77}, 321
(1976); A. Yoshizawa, {\it  Phys. Fluids} {\bf 30}, 1089 (1987);
M. Theobald, P. Fox, and S. Sofia, {\it Phys. Plasmas} {\bf 1}, 3016 (1994);
W-C. M\"uller and D. Carati, {\it Phys. Plasmas} {\bf 9}, 824 (2002).
B. Knaepen and P. Moin, {\it Phys. Fluids}, {\bf 16}, 1255, (2004).

\bibitem{ppp04}
Y. Ponty, H. Politano, and J.F. Pinton, {\it Phys. Rev. Lett.} {\bf 92}, 144503 (2004).

\bibitem{CL81}
J.P. Chollet and M. Lesieur, {\it J. Atmos. Sci.} {\bf 38}, 2747 (1981).

\bibitem{kmoins3}
A. Alemany et al., 
{\it J. M\'eca.} {\bf 18}, 277 (1979).

\bibitem{muller}
U. M\"uller, R. Stieglitz, and S. Horanyi, {\it J. Fluid Mech.}, {\bf 498}, 31 (2004)

\bibitem{petrelis}
F. P\'etr\'elis and S. Fauve, {\it Eur. Phys. J. B}, {\bf 22}, 273 (2001).
                                       
\end{thebibliography}
\end{document}